\input{psfig.sty}
\documentclass[12pt]{article}
\usepackage{graphicx}
\newfont{\feff}{cmti10}
\begin{document}

\title{Mean- Field Approximation and a Small Parameter in Turbulence Theory}

\author{ Victor Yakhot\\
Department of Aerospace and Mechanical Engineering\\
Boston University,
Boston, MA 02215 }

\maketitle

${\bf Abstract}.$
Numerical and physical experiments on two-dimensional ($2d$) turbulence
show that the differences of transverse components of velocity field
are well described by a gaussian statistics and Kolmogorov scaling exponents.
In this case the dissipation fluctuations are irrelevant in the limit of small 
viscosity. In general, one can assume existence of critical
space-dimensionality $d=d_{c}$, at which the energy flux and all odd-order
moments of velocity difference change sign and the dissipation fluctuations become
dynamically unimportant. At $d<d_{c}$  the flow can be described by the
``mean-field theory'', leading to the observed gaussian statistics and
Kolmogorov scaling of transverse velocity differences. It is shown that in the vicinity of $d=d_{c}$ 
the ratio of the relaxation and translation characteristic times decreases to zero, thus 
giving rise to a small parameter of the theory. 
The expressions for pressure and dissipation contributions
to the exact equation for the generating function of transverse velocity
differences are derived in the vicinity of $d=d_{c}$. The resulting equation
describes  experimental data on two-dimensional turbulence and demonstrate
onset of intermittency as $d-d_{c}>0$ and $r/L\rightarrow 0$ in
three-dimensional flows in close agreement with experimental data. 
In addition, some new exact relations between correlation 
functions of velocity differences are derived. It is also predicted that the 
single-point pdf of transverse velocity difference in developing as well as 
in the large-scale stabilized two-dimensional turbulence is a gaussian.
\title{Mean-Field Approximation and a Small Parameter in Turbulence Theory}

\author{ Victor Yakhot}

\newpage

\section{Introduction}

\noindent The role of the mean-field theories and gaussian limits 
as starting points for understanding of such important physical phenomena 
as superconductivity, superfluidity, critical point, 
naming just a few, can hardly be overestimated. These theories, usually based
on remarkable physical intuition and insight,  provided 
mathematical and intellectual basis for investigation of much more difficult
regimes in terms of  deviations from the mean field   solutions. 
The most recent example is a theory of anomalous
scaling in a model of a passive scalar,  advected by a random velocity field,
which was
developed as an  expansion in powers of small parameters  characterizing 
deviations  from  the two gaussian limits [1]-[3]. 
In a typical
non-linear system,  a gaussian limit corresponds  to a  weak
 coupling asymptotics  and, 
as a consequence,  to  a ``normal'',   non-anomalous, 
 scaling,  which can
often be obtained from a  ``bare' or linearized problem. A good example of
this behaviour is a fluid in thermodynamic equilibrium. 

\noindent The   large- Reynolds- number three-dimensional ($3d$) 
strong turbulence is 
characterized by an $O(1)$  energy flux ${\cal E}=\nu\overline{(\partial_{i}
v_{j})^{2}}$,  which in many flows is $O(v_{rms}^{3}/L)$, where $L$ is an
integral scale of turbulence. In the inertial range, where 
$k_{d}>>k\rightarrow \infty$ or $r/L<<1$, the observed energy spectrum $E(k)$ 
is close to  the one, 
proposed by Kolmogorov and the probability density $P(\Delta u)$ with 
$\Delta u=u(x+r)-u(x)$, is far from
the gaussian. Moreover, the experiments revealed that the moments of velocity
difference $S_{n,0}=\overline{(\Delta u)^{n}}
\propto r^{\xi_{n}}$ with the exponents $\xi_{n}$ 
given by very strange (``anomalous'') numbers,  which cannot be obtained on
dimensional grounds. This anomalous scaling and 
the very existence of the energy flux, resulting
in the non-zero value of  the third-order moment 
$S_{3,0}=\overline{(\Delta u)^{3}}\approx O(r)$, where ${\bf u\cdot r}=ur$,
implies a strongly non-gaussian process and an obvious 
lack of the mean-field limit. 

\noindent The situation may not be so grim, however: all odd-order moments 
of transverse velocity differences in both $2d$ and $3d$ flows 
$S_{0,2n+1}=\overline{(v(x+r)-v(x))^{2n+1}}=0$ with ${\bf v\cdot r=0}$.
This fact tells us that these
components of velocity differences do not participate in the inter-scale
energy transfer and there is no a'priori  any reason for them
not to obey  
gaussian statistics in some limiting cases. This is indeed true in
two-dimensional turbulence in the inverse cascade range where $l_{f}<<r<<L$
and $l_{f}$ is a forcing scale. 

\noindent Numerical and physical experiments on the  external- force- driven 
two- dimensional turbulence showed that the moments of 
transverse velocity differences  and
even-order moments of the longitudinal ones are very close to the gaussian
values and are characterized by the Kolmogorov scaling exponents
$S_{0,2n}\propto S_{2n,0}\propto r^{\xi_{2n}}$ with $\xi_{2n}=\frac{2n}{3}$
[4]-[7]. The odd-order moments of longitudinal velocity differences are positive
in a $2d$ flow, while they all are negative in $3d$. 
This observation tells that  the  most important distinction between two and 
three-dimensional turbulence in in 
dynamic  role of the dissipation contributions: they are irrelevant in the  
two-dimensional inverse cascade range and are crucial (see below) for the
small-scale dynamics of a
three-dimensional flow where the forcing terms can be neglected (see below).
 Thus, we can assume that it is the dissipation
fluctuations that are responsible for both strong deviations from the gaussian
statistics and anomalous scaling in three dimensional flows. This assumption
is consistent with the observation of the close-to-gaussian 
probability density of velocity differences in
$3d$ turbulence at the scales $r\approx L$ where 
the integral scale $L$ is defined as the one at which: $S_{3,0}(L)=0$  
and the intermittent 
 dissipation fluctuations disappear [8].  It is clear that this range
($r\approx L$) is not characterized by a well-defined scaling exponents.

\noindent As will become clear below, the forcing contribution to
 the equation for the probability density involves factor $\mu\approx 1-cos(k_{f}r)$. This means that 
at the scales $r>>1/k_{f}$ the parameter $\mu\approx 1$ while $\mu\approx (k_{f}r)^{2}$ when $k_{f}r<<1$. Thus, the forcing term must be important in the inertial range of two-dimensional turbulence ($r>>l_{f}$) and is irrelevant in 
the $3d$-inertial range with the positive energy flux. 
This change  happens at some
 space dimensionality $d=d_{c}$ at which the energy flux changes sign. 
First calculation of $d_{c}\approx 2.05$ was conducted by Frisch et. al. [9]
 within the framework of a simple closure model. The more physically 
transparent calculation can be performed for the  
Navier-Stokes equations driven by a random force having an 
algebraically decaying spectrum in the inertial range [10], where 
a one- loop
 small-scale- elimination procedure gives a correction to the 
bare viscosity:

$$\delta \nu=\nu\frac{d^{2}-d-\epsilon}{2d(d+2)}Re^{2}$$   

\noindent where $Re$ is a properly defined
 Reynolds number corresponding to the
 eliminated small-scale velocity fluctuations 
and $\epsilon$ is a parameter characterizing 
the forcing function. In case of Kolmogorov turbulence 
$\epsilon\approx 4$. This relation shows that the role, 
the small-scales play in turbulence dynamics,  depends on the space
 dimensionality $d$: the correction
 to viscosity is positive
 when $d>d_{c}(\epsilon)$ where
 it changes sign. Physically,  this means that
 the small-scale velocity fluctuations take energy
 from the large- scales motions (direct energy cascade
 from large-to-small structures) at $d>d_{c}$,  while at $d<d_{c}$ they 
excite the large scale motions
 (spend  their energy) giving rise to the inverse energy cascade.
 For $\epsilon=4$ the critical  dimensionality $d_{c}\approx 2.56$.
The correct value of $d_{c}$ is not too important: 
what is crucial for the theory presented below is 
that the critical dimensionality, 
 at which the flux changes its sign, exists. 
It will be shown below that $d-d_{c}\rightarrow 0$ is a small parameter of the theory enabling one to calculate an
expression for the dissipation anomaly in the form, resembling the Kolmogorov refined similarity
hypothesis.

\noindent Since in $2d$ the moments $S_{0,2n}$ show  Kolmogorov scaling,  
the  gaussian statistics of transverse velocity differences cannot 
correspond to the weak coupling limit. This problem was considered in the
Ref. [11] where, following Polyakov [12], 
the equation for the generating function 
for the problem of the  Navier-Stokes 
turbulence was first introduced . An unusual symmetry of this
 equation enabled one to show that the
solution was  consistent with both Kolmogorov scaling and gaussian
statistics. 
In this work a more detailed 
  theory of two-dimensional turbulence is
presented and the generalization to  three-dimensional flows is
considered. The main result of the paper is a model demonstrating how the
deviations from the ``normal scaling'' and gaussian statistics appear in $3d$
when the strength of the dissipation term $\beta >0$ 
and the ``scale'' parameter $\epsilon=1-r/L$
deviate from zero. 

\noindent This paper is organized as follows. In Section 2  the  equation
for the generating function, derived in [11] is introduced. Some new exact 
relations
between velocity structure functions, following from this equation, are
derived in Section  3. The connection between scaling exponents of the moments
of velocity differences and their amplitudes is established in  Section 4. 
The mean-field derivation of 
 the pressure term is given  in Section 5 which is
used to obtain  a  gaussian pdf in the two-dimensional flow in Section 6. 
In Sections  7 the small parameter of the theory is identified and used 
for derivation of  the expression for the dissipation contributions to the eqaution for the pdf.
Section 8 is devoted to solution of the equation and demonstration how 
anomalous scaling and deviations from gaussian statistics emergr from the theory.
Conclusive remarks are presented in Section 9.

\noindent 

\section{Equation  for the Generating Function}

\noindent 
The equations of motion are (density $\rho\equiv 1$):

\begin{equation}
\partial_{t}v_{i}+v_{j}\partial_{j}v_{i}=-\partial_{i}p+\nu\nabla^{2}v_{i}+f_{i};
~~~~~~\partial_{i}v_{i}=0
\end{equation}

\noindent where $\bf f$ is a forcing function responsible for  the 
large-scale kinetic energy  production  and in a statistically 
steady state the mean pumping rate $P
=\overline{{\bf f\cdot v}}$. 
In what follows we will be mainly interested 
in the probability density function of  two-point velocity
difference ${\bf U}={\bf u(x')-u(x)}\equiv \Delta {\bf u}$. The generating 
function is:
$Z=<exp{(\bf{\lambda\cdot U})}>$.  The equation for the generating function of
velocity differences corresponding to (1) is:

\begin{equation}
\frac{\partial Z}{\partial t}+\frac{\partial^{2} Z
}{\partial \lambda_{\mu}\partial  r_{\mu}}=I_{f}+I_{p}+D
\end{equation}

\noindent with 

\begin{equation}
I_{f}=<{\bf \lambda \cdot \Delta f}e^{\lambda\cdot \Delta {\bf u}}>
\end{equation}

\begin{equation}
I_{p} =
 -{\bf \lambda \cdot}<e^{\lambda\cdot \Delta {\bf u}}\Delta (\nabla p)>\equiv
-\lambda\cdot<
e^{{\bf \lambda\cdot v}}  
({\bf \nabla_{2}}p(x_{2})- {\bf \nabla_{1}}p(x_{1}))>
\end{equation}

\noindent and

\begin{equation}
D = \nu{\bf \lambda\cdot <(\nabla_{2}^{2} v(x_{2})-\nabla_{1}^{2} v(x_{1}))}
e^{\bf \lambda\cdot U}>
\end{equation}

\noindent The most interesting and surprising 
feature of (2) is the fact that, unlike in the problem 
 of Burgers
turbulence [12],  the
advective contributions are represented there  in a closed form. 
This means that the theory, developed below, is free from the
 troubles related to Galileo invariance, 
haunting
 all schemes, based on renormalized 
 perturbation expansions in powers of Reynolds
 number. To 
completely
close the problem the expressions  for $I_{p}$ and $D$  are  needed. The
equations (2)-(3) formulate the turbulence theory in terms of ``only'' two
unknowns $I_{p}$ and $D$. The Kolmogorov refined similarity hypothesis
stating that $(\Delta u)^{3}=\phi{\cal E}_{r}r$
where  $\phi$ is a scale-independent random process and
${\cal E}_{r}$ is a dissipation rate averaged over a ball of
radius $r$ around point $x$, can be a promising starting point to  a
closure for the dissipation term $D$. This will be done below. The pressure
term  in (2)-(3) is  also of a very specific and rather limited nature: 
all we have to
know is the correlation functions $<U_{i}U_{j}\cdot\cdot\cdot U_{m}\Delta {\bf
\nabla} p>$. Thus, the definite targets needed for derivation of the
closed equation for $Z$-functions are well-defined. 

The generating function can depend only on three variables:
$\eta_{1}=r;~~ \eta_{2}=\frac{{\bf \lambda\cdot r}}{r}\equiv 
\lambda cos(\theta);~~ \eta_{3}=\sqrt{\lambda^{2}-\eta_{2}^{2}};$ and   

\begin{equation}
Z_{t}+[\partial_{\eta_{1}}\partial_{\eta_{2}}+\frac{d-1}{r}\partial_{\eta_{2}}
+\frac{\eta_{3}}{r}\partial_{\eta_{2}}\partial_{\eta_{3}}+\frac{(2-d)\eta_{2}}{r\eta_{3}}\partial_{\eta_{3}}-\frac{\eta_{2}}{r}\partial^{2}_{\eta_{3}}]Z=
I_{f}+I_{p}+D
\end{equation}

\noindent where, to simplify notation we set $\partial_{i,\alpha}\equiv
\frac{\partial}{\partial x.\alpha}$ and $v(i)\equiv v({\bf x_{i}})$. 
Below we will often use $\partial_{\eta_{i}}\equiv \partial_{i}$.  The
functions $I_{p}$, $I_{f}$ and $D$ are easily extracted from the above 
definitions. Let us denote $\Delta u\equiv U$ and $\Delta v\equiv V$.

\noindent In the new variables the generating function can be 
represented as:

$$
Z=<e^{\eta_{2}\Delta u +\eta_{3} \Delta v}>\equiv  <e^{\eta_{2}U+\eta_{3}V}>
$$

\noindent  with the mean dissipation rate ${\cal E}$ defined by:
$\overline{\nu(\partial_{x_i} u)^{2}}=\frac{1}{d}{\cal E}$. Any correlation
function is thus:

$$
S_{n,m}\equiv 
<U^{n}V^{m}>=\partial^{n}_{2}\partial^{m}_{3}Z(\eta_{2}=\eta_{3}=0,r)
$$

\section{Relations between moments of velocity difference}

\noindent Let us discuss some direct consequences of  equations (1)-(6).
The Navier-Stokes equations are invariant under  transformation:
$v\rightarrow -v$ and $y\rightarrow -y$. That is why: 
$<(\partial_{y}p(0)-\partial_{y} p(r))(\Delta v)^{m}>\neq 0$ if 
$m=2n+1$ with $n>1$ 
and is equal to zero if $m=2n$. It is also clear from the symmetry that 
$<\partial_{x}p U^{2n}>=0$. 
It follows from the Navier-Stokes equations that 
$<(\nabla^{2} U)  U^{2n}>=<(\nabla^{2} V) V^{2n}>=0$.

\noindent 
Multiplying (6) by $\eta_{3}$, applying $\partial_{2}^{2n-1}$ to the resulting 
equation gives as $\eta_{2}=\eta_{3}\rightarrow 0$

\begin{equation}
\frac{\partial S_{2n,0}}{\partial r}+\frac{d-1}{r}S_{2n,0}-
\frac{(d-1)(2n-1)}{r}S_{2n-2,2}=
P(1-cos(r/l_{f}))\frac{2(2n-1)(2n-2)}{d}S_{2n-3,0}
\end{equation}

\noindent The right side of (7) is $O(r^{2})$ 
in a three-dimensional flow (3d)  where $r/l_{f}<<1$ and and can be
neglected. It is however 
is $O(1)$ in two dimensional turbulence in the 
inverse cascade range where  $r/l_{f}>>1$. The dissipation 
terms
do not contribute to this relation. 
For $n=1$ (7) gives a well-known incompressibility relation [13], [14]:

\begin{equation}
\frac{\partial S_{2,0}}{\partial r}+\frac{d-1}{r}S_{2,0}=\frac{d-1}{r}
S_{0,2}
\end{equation}

\noindent Multiplying (6) by $\eta_{3}$ with subsequent
$\partial_{3}\partial^{2}_{2}$ leads after setting
$\eta_{2}=\eta_{3}=0$ as $\nu\rightarrow 0$ to:

\begin{equation}
\frac{\partial S_{3,0}}{\partial r}+\frac{d-1}{r}S_{3,0}-2\frac{d-1}{r}S_{1,2}=
(-1)^{d}\frac{4}{d}P
\end{equation}

\noindent where $d=2;~3$ .
Applying $\partial^{3}_{3}\eta_{3}$ to (6)
gives,  as $\nu\rightarrow 0$:

\begin{equation}
\frac{\partial S_{1,2}}{\partial r}+\frac{d+1}{r}S_{1,2}=
(-1)^{d}\frac{4}{d}P
\end{equation}

\noindent Substituting this into (9) yields a well-known Kolmogorov relation:

\begin{equation}
S_{3,0}\equiv \overline {(\Delta u)^{3}}=(-1)^{d}\frac{12}{d(d+2)}Pr
\end{equation}

\noindent For $2n=4$ the relation (7) reads:

\begin{equation}
\frac{\partial S_{4,0}}{\partial r}+\frac{d-1}{r}S_{4,0}=\frac{3(d-1)}{r}S_{2,2}
\end{equation}

\noindent In two-dimensional turbulence $(d=2)$ 
in the inverse cascade range one can
neglect the dissipation contribution $D$ (see below) and derive:

\begin{equation}
\frac{\partial S_{1,2n}}{\partial r}+\frac{1+2n}{r}S_{1,2n}=
n(2n-1)PS_{0,2n-2}-2n<{\cal P}_{yv}(\Delta v)^{2n-1}>
\end{equation}

\noindent where ${\cal P}_{yv}\equiv{\partial_{y}v(x+r)-\partial_{y}v(x)}$. 
Another interesting relation, valid in $2d$ is obtained from (6) readily:

\begin{equation}
\frac{\partial S_{2,2n}}{\partial r}+\frac{1+2n}{r}S_{2,2n}=
\frac{S_{0,2n+2}}{r}-2n<{\cal P}_{yv} \Delta u (\Delta v)^{2n-1}>+
n(2n-1)PS_{1,2n-2}
\end{equation}

\noindent In the direct cascade range, where the forcing contribution is
$O(r^{2})\rightarrow 0$,  the relation (14) for an arbitrary
dimensionality $d$ reads:

\begin{equation}
\frac{\partial S_{2,2n}}{\partial r}+\frac{d-1+2n}{r}S_{2,2n}=
\frac{2n+d-1}{2n+1}\frac{S_{0,2n+2}}{r}-
2n<{\cal P}_{yv}\Delta u (\Delta v)^{2n-1}>
\end{equation}

\noindent Now, let us multiply (6) by $\eta_{3}$,
differentiate once over $\eta_{2}$ and three times over $\eta_{3}$. This
gives:

\begin{equation}
\frac{\partial S_{2,2}}{\partial r}+\frac{d+1}{r}S_{2,2}=
 \frac{d+1}{3r}S_{0,4}-2\overline
{{\cal P}_{yv}\Delta u \Delta v}
\end{equation}

\noindent This relation is correct since 
$\nu\overline{\nabla^{2}v\Delta u\Delta v}=
\nu\overline{\nabla^{2}u (\Delta v)^{2}}=0$.

\noindent {\bf 2d simulations of Bofetta, Celani and Vergassola} 
To achieve a true steady state these authors [7] 
conducted a series of 
 very accurate simulations of the problem (1) with the large-scale dissipation
term $D_{L}=-\alpha {\bf v}$ in the right side (6).  The 
moments of
transverse velocity differences, reported in this paper ($n\geq 2$),  
were very close to
their gaussian values. 
It is clear that 
this term introduces $-\alpha \lambda_{\mu}\frac{\partial Z}{\partial
\lambda_{\mu}}$ into the right side of (6) which is small in the inertial
range where the non-linearity is large. One has to be careful though
with the 
dangerous 
interval
$\Delta v\rightarrow 0$ where the linear term is not small. We  expect the 
negative-
order structure functions with $-1<n<0$  strongly depend on the functional
shape of the otherwise irrelevant large-scale dissipation term. The same can 
be
predicted for various 
 conditional expectation values of dynamical variables,  like
pressure gradients and dissipation terms,
 for the fixed values of $\Delta v \Delta u$: near the origin where 
$\Delta v$ and $\Delta u$ are very
small,
the artificially introduced 
linear contributions  to the Navier-Stokes equations
 dominates, producing large and non-universal 
 deviations from the universal functions characterizing inertial range. 
For example, with addition of the linear dissipation,  the relation (14) 
reads:

\begin{equation}
\frac{\partial S_{2,2n}}{\partial r}+\frac{1+2n}{r}S_{2,2n}=
\frac{S_{0,2n+2}}{r}-(2n+1)\alpha S_{1,2n}-
2n<{\cal P}_{yv} \Delta u (\Delta v)^{2n-1}>+
n(2n-1)PS_{1,2n-2}
\end{equation}

\noindent modifying the balance (pressure contribution) 
in the range of small product $\Delta u\Delta v$ or $r/L\approx 1$. 
In the interval where  $\Delta u \Delta v$ is  not small ($r\rightarrow 0$), 
the linear terms are
small and can be neglected. 

\noindent The results obtained in this section can also be derived with 

\begin{equation}
Z=e^{{\sqrt{d-1}\eta_{3}V+\eta_{2}U}}
\end{equation}

\noindent with the properly defined moments of $V$ and $U$.

\section{Asymptotic values of exponents in three-dimensional flows}

\noindent In case of intermittent turbulence $S_{m,n}=A_{m,n} r^{\xi_{m,n}}$.
We can see that in the inertial range of a three-dimensional flow
 ($r/l_{f}<<1$) the right side of (7) is negligible
and, as a result,  
$\xi_{2n,0}=\xi_{2n-2,2}\equiv \xi_{2n}<\frac{2n}{3}$. Substituting this
into (7) gives immediately:

\begin{equation}
\xi_{2n}=(d-1)[(2n-1)\frac{A_{2n-2,2}}{A_{2n,0}}-1]
\end{equation}

\noindent Let us introduce the probability density functions:

\begin{equation}
S_{2n,0}=\overline{U^{2n}}=\int P(U)U^{2n}~dU
\end{equation}

\noindent and 
\begin{equation}
S_{2n-2,2}=\overline{U^{2n-2}V^{2}}=\int P(U)U^{2n-2}q(V|U)dUdV
\end{equation}

\noindent where $q(V|U)$ is conditional pdf of $V$ for fixed value of $U$. 
It is clear that $q(V|U)=q(-V,U)$, so that all odd-order moments of $V$ are
equal to zero..
This expression can be rewritten as:

\noindent 
\begin{equation}
S_{2n-2,2}=\overline{U^{2n-2}V^{2}}=\int P(U)U^{2n-2}Q_{2}(U)dU
\end{equation}

\noindent where $Q_{2}$ is a conditional expectation value of $V^{2}$ for a
fixed value of $U$:

\begin{equation}
Q_{2}(U)=\int V^{2}q(V|U)~dV
\end{equation}

\noindent  Comparing the above expressions we observe that 
the amplitudes $A_{2n,0}\approx A_{2n-2,2}$ only if $Q_{2}\propto U^{2}$.
This seems rather improbable  in the limit $U\rightarrow \infty$
 $(n\rightarrow \infty)$. 
As a result,
the linear regime $\xi_{2n}\propto n$ is equally improbable. 

\noindent Saturation of exponents $\xi_{2n}\rightarrow \xi_{\infty}=const$ as 
$n\rightarrow \infty$ is
possible on a rather wide class of probability densities. For  example, 

\begin{equation}
P(U)\rightarrow -A\frac{1}{U}\frac{\partial}{\partial U} P(U)Q_{2}(U)
\end{equation}

\noindent where $A>0$ is a constant. Then, 
\begin{equation}
S_{2n,0}=(2n-1)A\int P(U)Q_{2}(U)U^{2n-2} dU=(2n-1)AS_{2n-2,2}
\end{equation}

\noindent Substituting this result into the expression for $\xi_{2n}$ gives
$\xi_{2n}\rightarrow (d-1)(A-1)=const$. The relation (24) defines the 
large-$U$
asymptotics of the pdf
$P(U)$ in terms of $Q_{2}(U)$:

\begin{equation}
P(U)\propto \frac{1}{Q_{2}(U)}e^{-A\int^{U}\frac{udu}{Q_{2}(u)}}
\end{equation}

\noindent If $Q_{2}(U)\rightarrow U^{\beta}$, then, assuming the existence of
all moments,  it follows from (25) that $\beta
< 2$.  The ``lognormal `` pdf $P(U)$ corresponds to $Q_{2}(U)\propto
U^{2}/2log(U)$. The expression (24)  also gives   in the  limit of large $n$:

\begin{equation}
S_{2n+1,0}=\overline{U^{2n+1}}=2nA\overline{V^{2}U^{2n-1}}=2nA S_{2n-1,2}
\end{equation}

\noindent

\section{Pressure contributions}

\noindent Due to the symmetries of the Navier-Stokes equations, 
neither  pressure nor dissipation terms contributed 
to the expressions (7) -(12). To proceed further we have to evaluate 
$I_{p}$ and $D$.

\noindent First of all we see from (12) that $\xi_{4,0}=\xi_{2,2}$. 
Let us assume that in the inertial range $S_{4,0}=A_{4,0}r^{\xi_{4,0}}$,
$S_{0,4}=A_{0,4}r^{\xi_{0,4}}$ and $S_{2,2}=A_{2,2}r^{\xi_{2,2}}$. 
Then, it is clear
from (12) and (16) that neglecting the pressure contribution to (16) gives
$\xi_{4,0}=\xi_{0,4}$. 

\noindent ${\bf d=2}$. It will be shown below that in 2d
the even-order moments of velocity
differences are very close to the gaussian ones and all exponents are close
 to the K41 values $\xi_{n}=n/3$.  Then, $A_{4,0}=3A_{2,0}^{2}$ and
$A_{0,4}=3A_{0,2}^{2}$. It follows from  (12) that   
$A_{2,2}=\frac{7}{9}A_{4,0}$. Taking into account that when $d=2$ the
amplitudes 
$\frac{5}{3}A_{2,0}=A_{0,2}$,  we conclude that without the pressure
contribution the equations ( 12) and (16) are incompatible. 

\noindent Following [15]  we introduce a conditional 
expectation value of the
pressure gradient difference for a fixed value of $\Delta u$, 
$ \Delta v$ and $r$:

\begin{equation}
<\partial_{y}p(x+r)-\partial_{y}p(x)|\Delta u, \Delta v, r>\approx 
\sum_{m,n}\kappa_{m,n}(r)(\Delta u)^{m}(\Delta v)^{n}
\end{equation}

\noindent where the functions $\kappa_{m,n}(r)$ ensure  
proper dimensionality of the
corresponding correlation functions.
The above expression explicitly assumes existence of an expansion of conditional expectation 
value (28). In general, this may not be true due to various singularities 
such as the ones arising in the dissipation contributions (see below). Since the pressure term
involves only one spacial derivative, the ultra-violet singularity cannot appear. The infra-red singularity 
is not there at least in 2d where the integral scale is time-dependent (see below).
Keeping only the first two terms  of the expansion
(28), produces  a model for the pressure contributions:
 
\begin{equation}
<\partial_{y}p(x+r)-\partial_{y}p(x)|\Delta y\Delta v>\approx -h\frac{\Delta
u\Delta v}{r}-b\frac{\Delta v}{(Pr)^{\frac{2}{3}}}
\end{equation}

\noindent  Since in an  incompressible and homogeneous flow 
$\overline{\Delta u  {\cal P}_{xu}}=\overline{\Delta v  {\cal P}_{yv}}=0$, the
coefficients $h$ and $b$ are related as:

\begin{equation}
-hS_{1,2}=bS_{0,2}(Pr)^{\frac{1}{3}}
\end{equation}

\noindent  Limiting the expansion of a  conditional expectation value by
the first terms   resembles
Landau's  theory of critical phenomena, well describing experimental data in a
certain range of parameters variation.  We will show below that in case of
turbulence this approximation gives the 
results which are in agreement with the
data. This may be a consequence  of the fact that $A_{0,4}\approx 
A_{4,0}\approx
A_{2,2}=O(1)$.

\noindent With $\xi_{n}=n/3$ it follows from (7), (13)-(15)  and  (29) 
that when $d=2$ and $n\rightarrow \infty$:

$$
(\frac{2n(4-3h)}{3}+1)\frac{S_{1,2n}}{r}=n(2n-1)PS_{0,2n-2}+
b\frac{2nS_{0,2n}}{(Pr)^{\frac{2}{3}}}$$

\noindent and 

$$
(\frac{2n(4-3h)}{3}+1)\frac{S_{2,2n}}{r}=
\frac{S_{0,2n+2}}{r}+n(2n-1)PS_{1,2n-2}+b\frac{2nS_{1,2n}}{(Pr)^{\frac{2}{3}}}
$$

\noindent In the limit $n\rightarrow \infty$ assuming that $S_{1,2n}
\approx nS_{1,2n-2}$ one derives readily:

\begin{equation}
PrnS_{0,2n-2}\approx \frac{S_{2,2n+2}}{Prn}\approx \frac{1}{Prn^{2}}S_{0,2n+4}
\end{equation}

\noindent which is 
consistent with the gaussian pdf $P(\Delta v)$ as $\Delta v
\rightarrow \infty$. Thus, the relation 
(29) implies  
the gaussian tails of the probability density. It is clear
that due to the finite energy flux and relations (10)-(11), 
two-dimensional turbulence cannot be a gaussian process. All the relation (31)
can tell us is that the even-order moments with $n>>1$, described by (31),
 can
be close to the gaussian values. It will be shown below that transverse
velocity differences, not directly involved in the inter-scale 
energy transfer,  can obey  gaussian statistics.

\noindent It is clear from (17) that the model (29) for the  pressure
contributions is wrong  when the linear dissipation terms are added to
the Navier-Stokes equations.  In the limit of small $\Delta u\Delta v$ the
balance is achieved when:

$$
<{\cal P}_{yv}|\Delta u,\Delta v,r>\approx -h\frac{\Delta u\Delta
v}{r}-(\alpha+\frac{b}{(Pr)^{\frac{2}{3}}})\Delta v
$$

\noindent which differs from (29) in the range of small $\Delta u\Delta v$.

\noindent ${\bf 3d}$. In the intermittent 
three-dimensional turbulence
$\xi_{2n}<\xi_{2n+1}$. This produces strong restrictions 
on  the structure  of the
pressure contributions to the equation (6). Let us assume that
$\xi_{2n,0}=\xi_{2,2n-2}=\xi_{2n-2,2}$. Then, it is clear from (15) that
the first term of expansion (28) has all right properties.  The 
relation (15),  involving the  $r$-derivatives, 
  is valid for an arbitrary value of $n$ and that is why 
any additional term
of  expansion (28) 
must not only depend on a proper power of $n$ but  the
functions $\kappa_{m,n}(r)$ must also
reflect non-trivial dimensionalities caused by the anomalous scaling exponents
$\xi_{n}$. It cannot be rigorously ruled out, though 
this possibility seems quite bizzarre.
In what follows we will adopt the pressure model (29) in the three-dimensional
case also.

\section{Two-dimensional turbulence}

Now we will be
interested in the case of the two-dimensional turbulence in the inverse
cascade range.
If
a two-dimensional (2d) fluid   is stirred by a random (or non-random) forcing,  acting on a scale 
$l_{f}=1/k_{f}$,  the produced  energy is spent on creation of the large-scale
 ($l>l_{f}$)
flow which cannot be dissipated in the limit of  large Reynolds number as
$\nu\rightarrow 0$. This  
is a direct
and most important consequence of an additional,  enstrophy conservation, 
 law,
characteristic of two dimensional hydrodynamics [16]. 
As a result, the dissipation terms are irrelevant
in the inverse cascade range and we set $D=0$ in (6) and 
hope that in two dimensions the situation is greatly simplified. This hope is
supported by recent numerical and physical experiments [4]-[7]
 showing that as long as
the integral scale $L_{i}\propto t^{\frac{3}{2}}$ is  much smaller than the
size of the system, the velocity field at the scales $L_{i}>>l>>l_{f}$ is a
stationary close-to-gaussian process characterized by the structure functions
 with the Kolmogorov exponents $\xi_{n}=n/3$. 
In a recent paper Bofetta, Celani and Vegrassola [7] reported the
results of very accurate numerical simulations of two-dimensional turbulence
generated by a  random force. No deviations from  gaussian statistics of transverse
velocity differences as well as from the Kolmogorov scaling $\xi_{n}=n/3$
were detected. 
and is equal to zero if $m=2n$. 

\noindent The pressure gradient
$\partial_{y}p=\partial_{y}\partial_{i}\partial_{j}\partial^{-2}
\Delta v_{i}\Delta v_{j}$
and the difficulty in calculation of $I_{p}$ is in 
the   integral over the
entire space defined by  the inverse Laplacian $\partial^{-2}$. 
The huge  simplification,  valid  in 2d, 
 comes from the fact that all contributions to 
the left side of equation (6) as well as $I_{f}$ 
are independent on time. This means that the
integrals involved in the pressure terms 
cannot be infra-red divergent since
in a two-dimensional flow $L=L(t)\propto t^{\frac{3}{2}}$. 
We also have that $i_{p}\rightarrow \alpha^{2}i_{p}$ when $U,~V~\rightarrow 
\alpha U;~\alpha V$. 
Based on this and taking into account that
$<(\Delta v)^{2n+1}(u_{x}^{2}+v_{y}^{2}+u_{y}v_{x})>=0$, 
we, in the limit $\eta_{2}\rightarrow 0$,  adopt a low-order 
 model (29) giving:

\begin{equation}
I_{p}=
[h\frac{\partial^{2}}{\partial \eta_{2}\partial \eta_{3}}+
b\frac{\eta_{3}}{(Pr)^{\frac{2}{3}}}\partial_{3}]
Z(\eta_{2}=0,\eta_{3},r)
\end{equation}

\noindent In two dimensions the relation (32) combined with (6) in the limit
$\eta_{2}\rightarrow 0$,  solves the problem 2d turbulence. 

\noindent Substituting (32), (29) 
into (6) and,  based on (9)-(11),  seeking a solution as 
$\eta_{2}\rightarrow 0$ as 

\begin{equation}
Z(\eta_{2},\eta_{3},r)\approx Z_{3}(\eta_{3},r)\varphi(\eta_{2}r^{\frac{1}{3}}, \eta_{2})\approx 
Z_{3}(\eta_{3},r)exp(\frac{1}{2}A_{2,0}(\eta_{2}Pr^{\frac{1}{3}})^{2})
(1+\frac{1}{2}A_{1,2}\eta_{3}^{2}\eta_{2}(Pr)+\frac{\eta_{2}^{3}}{4}+...)
\end{equation}

\noindent where ($A_{1,2}=1/2$), gives:

\begin{equation}
[\partial_{r}+\frac{1}{r}+
\frac{1-h}{r}\eta_{3}\partial_{3}]\frac{A_{1,2}Pr}{2}\eta_{3}^{2}Z_{3}=2P\eta_{3}^{2}Z_{3}+\frac{b\eta_{3}}{r^{\frac{2}{3}}}\partial_{3}Z_{3}
\end{equation}

\noindent Setting 
 $Z_{3}=Z_{3}(\eta_{3}r^{\frac{1}{3}})\equiv Z_{3}(X)$ and  $h=4/3$ 
one derives using,  the relation (30) ($b=-\frac{2}{3A_{0,2}}$):

$$2A_{0,2}XZ_{3}=\partial_{X}Z_{3}$$

\noindent corresponding to a  Gaussian  solution with the 
correct width $A_{0,2}$. This fact serves as a consistency check that the gaussian is a solution for the PDF of transverse velocity differences.

\noindent
The equation (34) defines a 
probability density function corresponding to the finite moments 
$S_{2m,2n}(r)$ only when $h=4/3$.
 This situation resembles Polyakov's theory of 
Burgers turbulence [12] reduced to an eigenvalue problem with 
a single eigenvalue 
corresponding to the pdf which is positive in the entire interval.  
  
\noindent Having these exact  results, and keeping in mind (33) 
one can integrate the  equation over $\eta_{2}$ 
from $-i\infty$ to $0$ 
to obtain :

\begin{equation}
\frac{\partial Z_{3}}{\partial r}+3(1-h-b)
\frac{\eta_{3}}{r}\frac{\partial Z_{3}}{\partial
\eta_{3}}=\frac{2P}{(Pr)^{\frac{1}{3}}}\eta_{3}^{2}Z_{3}
\end{equation}

\noindent valid as long as   

$$\frac{(\eta_{3}r^{\frac{1}{3}})^{2}}{8A_{2,0}^{2}}<<1$$

\noindent  This constraint is an artifact of an approximate relation (33).  As will be shown below (35) gives an exact gaussian solution and thus,  is valid beyond above interval. 
 This result is obtained 
choosing the integration function $\Psi(\eta_{3},r)$ 
to compensate the $O(Z/r)$
term violating the normalizability constraint $Z(0,0,r)=1$. Solution to (35)
 is:

\begin{equation}
Z_{3}=exp(\gamma\eta_{3}^{2}(Pr)^{\frac{2}{3}})
\end{equation}

\noindent with the parameter $\gamma=
\frac{3}{(3(1-h-b)+1)}\propto A_{0,2}$ defining the width of the
gaussian. 

\noindent The first-order differential 
 equation (35) for the generating function 
differences implies the 
underlying linear Langevin dynamics of transverse velocity
differences. It is important that this equation in non-local in physical space
but local in the Fourier one. The effective forcing, corresponding to 
the right-side of (35), 
is a non-local and solution- dependent.

\noindent To evaluate the single-point probability density, corresponding 
to velocity differences in the limit 
of large displacements $r$ , 
we can notice that the energy flux 
is not equal to zero only at $r<<L$. At the distances 
$r\geq L(t)$ the zero value of the energy 
flux and symmetrization of the probability density ($P(\Delta u,L)=P(-\Delta u,L)$ can be achieved only when
the pressure contribution to the equation (6) 
compensates  the advective terms. As a result, since $D=0$, we have:

$$Z_{t}=2P\eta_{3}^{2} Z$$

\noindent Seeking a solution as $Z=Z(\eta_{3}\sqrt{t})\equiv 
Z(\eta_{3}v_{rms}(t))$ gives a gaussian result:

$$Z=e^{\eta_{3}^{2}u_{rms}^{2}(t)}$$

\noindent Similar outcome  is obtained for the case investigated in [7]. When 
turbulence is stabilized at 
the large scales by an artificially introduced friction,
the resulting equation is:

$$\eta_{3}\partial_{3} Z=2p \eta_{3}^{2} Z$$

\noindent also leading to the gaussian pdf. To conclude this section we would like to discuss the physical meaning of the integral scale $L$. The integral scale of turbulence is a scale at which the flux 
decreases  to zero [18] and at which $S_{3,0}(L)=0$.

\section{ Small Parameter in Turbulence Theory in Three Dimensions}
Three-dimensional turbulence is a notoriously difficult problem
due to absence of the small parameter. This can be illustrated on an example 
of a coarse-graining procedure, which was extremely successful in
 engineering turbulence simulations. Consider the wave-number  $k$ in the inertial range. 
Let us denote ${\bf v^{<}(k)}$ the Fourier components of the velocity field 
with  all modes ${\bf v^{>}(q)}=0$ when  $q>k$. The coarse-grained field
in the physical space is defined then:

$${\bf v}_{r}({\bf x},t)= \int_{k<\frac{1}{r}}{\bf v^{<}(k)} e^{i{\bf k\cdot x}} d^{3}k$$

\noindent The equation of motion for $v_{i}^{<}(k)$ in the Fourier space  
resembles the Navier-Stokes equation with effective viscosity

$$\nu(k)\approx 
(\frac{(d-d_{c})(d+\frac{1}{2})}{3.2d(d+2)})^{\frac{1}{3}}
{\cal E}^{\frac{1}{3}}k^{-\frac{4}{3}}$$

\noindent with $d_{c}\approx 2.56$  plus high order non-linearities. The
parameter $3.2$ in the above relation, evaluated at $d=3$,  
 is in fact a weak function  of 
space dimensionality and ${\cal E}=P=O(1)$.  This expression is derived assuming a close-to-gaussian statistics 
of the small-scale turbulence with

$$\overline{v_{i}(k,\omega)v_{j}(k',\omega')}\propto \frac{k^{-d}}{-i\omega +\nu(k) k^{2}}\delta(k+k')\delta(\omega+\omega')$$

\noindent which is accurate at $d-d_{c}\rightarrow 0$ [10] (see below).

\noindent In the physical space the effective viscosity of the coarse-grained field: 

$$\nu_{r}\approx (d-d_{c})^{\frac{1}{3}}N ({\cal E} r^{4})^{\frac{1}{3}}\approx v_{r}^{2}\tau_{r}$$
defines the relaxation time $\tau_{r}$ which is a characteristic time of 
interaction of the field $v_{r}$ with the eliminated modes acting on the scales $l<r$.

\noindent The difficulty of the theory is in the higher non-linearities

$$(v_{r}\nabla)^{n}\tau_{r}^{n-1}v_{r}$$

\noindent   and the dimensinless expansion parameter

$$\frac{\tau_{r}}{\theta_{r}}\approx \nabla_{r} v_{r} \tau_{r}\approx \frac{\tau_{r}v_{r}}{r}$$

\noindent is nothing but the ratio of translational time $\theta_{r}$,  characterizing the tendency of the ``large-scale'' 
longitudinal 
velocity fluctuation at the scale $r$ to form a small-scale structure (``shock'') in the absence of pressure, to the relaxation time 
strongly influenced by the pressure gradients contributions. In both 2d and 3d these times are of the same order and that is why 
trancation of the  expansion is a very difficult problem.

\noindent This is not so in the vicinity of $d=d_{c}$. Let us assume that the theory can be analytically continued 
to the non-integer dimensions [9]. Then, since the energy flux is $O(1)$ we have:

$${\cal E}\approx \frac{\partial}{\partial r}<\Delta u (\Delta v)^{2}>\approx \frac{\partial}{\partial r}<(\Delta u)^{3}>\approx 
\nu <(\frac{\partial v_{ri}}{\partial r_{j}})^{2}>=O(1)  $$

\noindent This means that 

$$v_{r,rms}\approx (d-d_{c})^{-\frac{1}{6}}({\cal E}r)^{\frac{1}{3}}$$

\noindent the dissipation wave-number $k_{d}\approx (d-d_{c})^{\frac{1}{4}}(\frac{{\cal E}}{\nu_{0}^{3}})^{\frac{1}{4}}$ and,  
as a consequence,  

$$\frac{\tau_{r}}{\theta_{r}}\approx (d-d_{c})^{\frac{1}{2}}$$

\noindent which will serve as a small parameter of the theory when $d-d_{c}\rightarrow 0+$.

\noindent These relations tell us that the turbulent intensity grows to infinity with 
$d-d_{c}\rightarrow 0$ where the energy flux changes its sign. The time needed to reach the steady stae is estimated easily:

$$T\approx (d-d_{c})^{-\frac{1}{3}}{\cal E}^{-\frac{1}{3}}r^{\frac{2}{3}}$$

\noindent after which a close-to-Kolmogorov spectrum is expected both above and below $d_{c}$. Thus, at $d=d_{c}$ the 
the flow is unsteady. 

\noindent The above results enable one to derive a plausable estimate for the dissipation term $D$. 
It is clear from the Navier-Stokes equations that:

$${\cal E}=-\frac{1}{2}\partial_{i}v_{i}v_{j}^{2}-\frac{1}{2}v^{2}_{t}-\partial_{i}v_{i}p$$

\noindent The coarsed-grained expression in the law frequency limit

\begin{equation}
{\cal E}\approx -\frac{1}{2}\frac{\partial}{\partial r_{r}} v_{ri}v_{rj}^{2}(1+O(d-d_{c}))
\end{equation}
\noindent To arrive in the expression for $D$ we assume $v_{r}\approx \Delta v$ leading to an expression very similar to 
Kolmogorov's  refined similarity hypothesis.

\section{Three-Dimensional Flow}

\noindent The most important feature of two-dimensional turbulence,  considered
in a previous section, is irrelevance of the dissipation processes in the
inverse cascade range when $d<d_{c}$. It is this irrelevance 
which was responsible for the
gaussian probability density of
 transverse velocity differences.

\noindent The model for the dissipation contribution $D$ in the limit $\eta_{2}\rightarrow 0$
is evaluated from (37) readily. We would like to keep at least some information about $\Delta v$ and the expression must 
be invariant under transformation ${\bf v}\rightarrow {\bf -v}$ and ${\bf x}\rightarrow -{\bf x}$.  In addition,
 the expression must be local in  physical space. Based on this considerations:

$${\cal E}_{v}\approx  c(d)\Delta u \Delta v \frac{\partial \Delta v}{\partial r}$$ 

\noindent where ${\cal E}_{v}$  is a dissipation rate of the ``v-component to kitenic energy'' $K_{v}=\frac{1}{2}v^{2}$.
Locality of this model is clear since $\partial_{r} \Delta v=\partial_{1} v(x_{1}) +\partial_{2} v(x_{2})$. 
The problem is in evaluation of the coefficient $c(d)$ since, in principle, it can be singular at $d=d_{c}$.   
Indeed, it is clear that the dissipation term is zero at $d\leq d_{c}$. However, the point $d=d_{c}$ separating 
inverse and direct cascade ranges,  is a singularity due to infinitly
large  
amplitudes of velocity fluctuations. All we can say at this point is that the pdf can be represented as a sum of even and odd 
functions of $\Delta v$. The symmetric part has a growing with $d-d_{c}\rightarrow 0+$ width,  while the width 
of odd one is $O(1)$. The behaviour of $c(d)$ in the vicinity of $d_{c}$ is not clear. We feel that it is $O(1)$ at $d>d_{c}$ and 
zero at $d\leq d_{c}$.  

\noindent Thus, 
we have:

\begin{equation}
D \approx 
 c(d)\eta_{3}\partial_{\eta_{2}}\partial_{\eta_{3}}\partial_{r}Z \nonumber \\
\approx 
c(d)\eta_{3}^{2}<\Delta u\Delta v\frac{\partial \Delta v}{\partial r} e^{\eta_{2}\Delta
 u+\eta_{3}\Delta v}>+O(\partial_{r}\partial_{\eta_{2}} Z)
\end{equation}

\noindent 
This expression obeys the 
basic symmetries of the Navier-Stokes equation.
The expression, similar to (38), was being used in engineering turbulent
modelling based on the low-order in the Reynolds number 
coarse-grained equations.

The last term in the right side of (38), simply modifies
the coefficient in front  of the first term in the left side 
the equation (6) and does not generate anything new.
The expression (38)
 resembles Kolmogorov's
refined similarity hypothesis,  connecting the dissipation rate, 
averaged over a
 region of radius $r$,  with $(\Delta u)^{3}$. 
Thus,  in the limit $\eta_{2}\rightarrow 0$:

\begin{equation}
[\partial_{\eta_{1}}\partial_{\eta_{2}}+\frac{2}{r}\partial_{\eta_{2}}+
(1+h)\frac{\eta_{3}}{r}\frac{\partial^{2}}{\partial_{\eta_{2}}\partial{\eta_{3}}}+
c(d)\eta_{3}\partial_{\eta_{2}}\partial_{\eta_{3}}\partial_{r}]                
Z(\eta_{2}=0,\eta_{3},r)=\nonumber \\
\frac{2P}{3}(1-cos(k_{f}r))Z -b\frac{\eta_{3}}{r^{\frac{2}{3}}}\partial_{3}Z
\end{equation}

\noindent 

The $d-d_{c}>0$  counterpart of equation (35) is:

\begin{equation}
[\partial_{\eta_{1}}+
(1+h+b)\frac{\eta_{3}}{r}\frac{\partial}{\partial{\eta_{3}}}+ 
c(d)\eta_{3}\partial_{\eta_{3}}\partial_{r}]                
Z(\eta_{2}=0,\eta_{3},r)=a\frac{2P}{3}\frac{1-cos(k_{f}r)}{r^{\frac{1}{3}}}\eta_{3}^{2}Z_{3}
\end{equation}

\noindent with $\Psi(\eta_{3})$ chosen in such a way that the generating
function $Z(0,0,r)=1$.
We consider  two limiting cases.

\noindent ${\bf  Small-Scale Dynamics }$.

\noindent 
Inverse Laplace transform of (40) without right side  
gives an  equation for the pdf $P(\Delta
v,r)$:

\begin{equation}
\frac{\partial P}{\partial
  r}+\frac{1+3\beta}{3r}\frac{\partial}{\partial V}VP
  -\beta\frac{\partial}{\partial V}V\frac{\partial P}{\partial r}=0
\end{equation}

\noindent where $\beta\propto c(d)$. Since $S_{0,3}=0$,  the 
coefficients in (41) are chosen 
to give $s_{0,3}=\overline{|\Delta v|^{3}}=a_{3} Pr$ with an undetermined
amplitude $a_{3}$. This is an assumption of the present theory, not based  on 
rigorous theoretical considerations. 
Seeking a solution in a form $S_{0,n}=<(\Delta v)^{n}>\propto 
r^{\xi_{n}}$ gives:

\begin{equation}
\xi_{n}=\frac{1+3\beta}{3(1+\beta n)}n\approx \frac{1.15}{3(1+0.05 n)}n
\end{equation}

\noindent which was derived in [18] together  with $\beta\approx 0.05$. 
It follows from (42) that:
$P(0,r)\propto r^{-\kappa}$
where $\kappa= \frac{1+3\beta}{3(1-\beta)}\approx 0.4$ for
$\beta=0.05$. Very often the experimental data are presented as $P(X,r)$ where 
$X=V/r^{\mu}$ with $2\mu=\xi_{2}\approx
0.696$ for $\beta=0.05$. This gives  
$P(X=0,r)\propto r^{-\kappa +\mu}\approx r^{-0.052}$
compared with the experimental data by Sreenivassan [19]: 
 $-\kappa +\mu \approx -0.06$. 

Let us write 
$P(V,r)=r^{-\kappa}F(\frac{V}{r^{\kappa}},r)=
r^{-\kappa}F(Y,r)$, so that $F$ obeys the  following equation:

\begin{equation}
(1-\beta)r\frac{\partial F}{\partial r}+\beta\kappa\frac{\partial}{\partial
Y}Y^{2}\frac{\partial F}{\partial Y}-\beta Yr\frac{\partial^{2} F}{\partial
Y \partial r}=0
\end{equation}

Next, changing the variables again 
$-\infty<y=Ln (Y)<\infty$, substituting this into (43) and 
evaluating the
  Fourier
 transform of
  the resulting 
  equation gives:

\begin{equation}
(1-\beta)r\frac{\partial F}{\partial r}+\beta\kappa
(ik-k^{2})F-ik\beta r\frac{\partial F}{\partial r}=0
\end{equation}

\noindent with the result:
$F\propto r^{\gamma(k)}$, 
where  
$\gamma(k)=\beta\kappa\frac{-ik+k^{2}}{1-\beta-i\beta k}Ln(r/L)$ 
with $r/L<<1$. We have to evaluate the inverse Fourier transform:

\begin{equation}
F=\int_{-\infty}^\infty dk e^{-iky}e^{\gamma(k)}
\end{equation}

\noindent in the limit $y=O(1)$ and $r\rightarrow
0$ so that $Ln(r/L)\rightarrow -\infty$. The integral can be calculated 
exactly. However, the resulting expression is very involved. Expanding  the
denominator of $\gamma(k)$ gives
:

\begin{equation}
F=\int_{-\infty}^\infty dk 
e^{-ik(y+\frac{\beta\kappa (Ln (r))}{1-\beta})}
e^{-\frac{\beta\kappa (1+\beta)|Ln (r)|}{(1-\beta)}k^{2}}
\end{equation}

\noindent and 

\begin{equation}
F\propto \frac{1}{\sqrt{\Omega (r)}}exp{(-\frac{(Ln(\xi))^{2}}{4\Omega})}
\end{equation}

\noindent with
$\xi=V/r^{\frac{\kappa}{1-\beta}}$ 
and
$\Omega (r)=4\beta\kappa\frac{1+\beta}{1-\beta}|Ln(r/L)|$.

\noindent To understand the range of validity of this expression, let
us evaluate $<V^{n}>$ using the expression (47) for the pdf. Simple 
integration, neglecting $O(\beta^{2})$ contributions, gives: 
$<V^{n}>\propto r^{\alpha_{n}}$ ~with 
$\alpha_{n}=(1+3\beta)(n-\beta (n^{2}+2))/3$. Comparing this relation
with the exact result (43) we conclude that the expression for the pdf,
calculated above, is valid in the range $n>>1$ and $\beta n<<1$. The
properties of the pdf in the range $3\leq  \xi \leq 15$ are
demonstrated on Fig. 1 
 for $r/L=0.1;~0.01;~0.001$. 
The log-normal distribution (48),  is valid in a certain (wide but limited) 
range of the $V$- variation. It is clear from (42)
that neglecting the dissipation terms ($c(d-d_{c})\propto\beta =0$)
 leads to 
$\xi_{n}=n/3$,  i. e. 
 disappearance  of anomalous scaling of  moments of velocity
differences.  This result agrees with the well-developed phenomenology,
attributing intermittency to the dissipation rate fluctuations: the
stronger the fluctuations, the smaller the fraction of the total space
they occupy [8], [14]. To the best of our knowledge, this is the first work 
leading to multifractal distribution of velocity differences as a result of
approximations made directly on the Navier-Stokes equations.

\noindent To investigate the probability density $P(Y,r)$ in
the limit $Y\rightarrow 0$ we introduce an expansion:

\begin{equation}
F(Y,r)=\sum_{n}  C_{n}Y^{2n}f_{2n}(r)
\end{equation}

\noindent Substituting this into (44) gives:

\begin{equation}
f_{2n}\propto (\frac{r}{L})^{-\frac{\beta\kappa 2n(2n-1)}{1-\beta(1+n)}}
\end{equation}

\noindent It is seen from (48)-(49) that the pdf starts bending from the
log-normal slope (47)
 toward $\partial_{Y}F(Y,r)=0$ at $Y=0$ 
at:

\begin{equation}
Y~<~(\frac{r}{L})^{\frac{\beta\kappa}{1-2\beta}}
\end{equation}

\noindent This inequality shows
that as $r\rightarrow 0$ the pdf develops a narrow cusp at the origin $Y=0$.
If the probability density is plotted in the dimensionless variable $X$, the 
bending starts at $X\approx r^{0.07}$. This value was
calculated, as above, 
with $\beta=0.05$.  

\noindent {\bf Large-scale limit: $r/L\approx 1$}.
\noindent 
Now, let us investigate the large-scale 
limit $\frac{r}{L}\rightarrow 1$. Realizing that $\beta$ can be an $O(1)$ constant we, for the illustration purposes, 
will 
investigate the large scale limit pretending that $\beta(d)\rightarrow 0$ as $d\rightarrow d_{c}$. This is also useful since the estimated value of $\beta\approx 0.05$ at $d=3$ which is numerically small.
In this limit  
the right side of (41) 
is $O(\eta_{3}^{2}Z)$ and cannot be neglected. 
Repeating the procedure 
leads  to an equation:

\begin{equation}
\frac{\partial P}{\partial
  r}+\frac{1+3\beta}{3r}\frac{\partial}{\partial V}VP
  -\beta\frac{\partial}{\partial V}V\frac{\partial P}{\partial r}=
a\frac{P}{(\Pr)^{\frac{1}{3}}}\frac{\partial ^{2} Z}{\partial V^{2}}
\end{equation}

\noindent where $a$ is a proportionality coefficient and $r\approx L$.
As one
can see from this equation in the limit of small $\beta$ the solution to this
equation approaches gaussian. It is also clear that for any finite $\beta$,
the tails of the pdf are strongly non-gaussian when

\begin{equation}
\beta Y^{2}~>>~1
\end{equation}

\noindent This estimate means that,  according to the theory presented above, 
the perturbative  treatment of  deviations from the mean-field
gaussian theory is possible but it involves two parameters: the ratio
$\epsilon=1-\frac{r}{L}<<1$ and $\beta<<1$. The fact that the ``real-life''
$\beta \approx 1/20$ may explain why the experimentally observed pdf of the large-scale
($\epsilon<<1$ ) velocity fluctuations was  so close to the gaussian [see [8]
and  references therein].

\noindent It is also seen from (52) that at $(\frac{r}{L})^{2}\approx \beta
\approx 0.05$ the pdf is dominated by a gaussian central part.

\section{Conclusions}

\noindent The equation (6) formulates theory of turbulence in 
terms of
``only''
two unknowns: pressure and dissipation terms $I_{p}$ and $D$, respectively. 
It provides a mathematical testing ground for various analytic 
expressions and
models obtained from numerical simulations.   

Armed with  the 
experimental and
numerical data,  supporting  gaussian statistics of transverse velocity
differenced in two-dimensional flows,  we showed that the mean field
approximation (the lowest-order term of the expansion (28))
for the pressure contributions  (29) leads to both  Kolmogorov scaling and
gaussian statistics  of transverse velocity differences. In addition, the
equation (6) shows that the single-point pdf's 
 in $2d$  turbulence are gaussian.  It is to be stressed that $2d$ turbulence
cannot be a gaussian process and probability density $P(\Delta u, \Delta v,r)=$
is not a gaussian. It is only pdf $P(\Delta v,r)=\int P(\Delta u,\Delta
v,r)d\Delta u$ is a gaussian. This statement  violates no  dynamic
constraints.   

\noindent One of the most interesting outcomes of the present theory 
is a discovery of existence of the two time-scales in the system which are very different in the vicinity of $d=d_{c}$.
This difference enables one to coarse-grain the Navier-Stokes equations and neglect
 all high-order non-linearities, generated by the procedure.  
Using this result the model for
for the dissipation term $D$ was derived. 

\noindent The as yet unresolved ambiguity of this model is its behaviour as $d\rightarrow d_{c}$. If transition from 
3d to the non-intermittent  state at $d<d_{c}$ is smooth, then $\beta\rightarrow 0$ and the 
resulting equation shows onset of both 
anomalous scaling and non-gaussian statistics. The transition can be singular, however: right at $d>d_{c}$ the 
coefficient $\beta$ can become $O(1)$  and a weakly intermittent state and weak coupling limit do not exist. 
In this case, due to existence of the small parameter, enabling evaluation of the dissipation expression $D$,
 the theory non-perturbatively  predicts both the shape of the pdf and scaling exponents provided the small parameter 
$S_{3,0}/(S_{2,0})^{\frac{3}{2}}\rightarrow 0$ as $d\rightarrow d_{c}$. This result is possible since $D=D_{0}+O(d-d_{c})$
avd even in the limit $d\rightarrow d_{c}$, the model $D=O(1)$. At $d-d_{c}<0$ ,  $D=0$ leading to the gaussian pdf of
transverse velocity differences.
\noindent
Experimental investigation of hydrodynamics in a non-integer space dimension is impossible. 
However,  it was demonstarted by Jensen [20] that a force-driven shell model is yields the changing  
sign of the energy flux upon variation of 
a leading parameter.  Numerical solution at a critical point (zero flux) demonstrated an unsteady state with the growing
total energy and the energy spectrum concentrated in the vicinity of $k_{f}$. The calculat
ion also gave kolmogorov energy spectrum at 
$''d>d_{c}''$ with growing Kolmogorov constant as $d\rightarrow d_{c}$. It is not yet clear how the eddy viscosity approximation 
works for the shell model,  but, since the $O(1)$ energy flux is 
fixed by the forcing function,   the growth of kinetic energy must be 
 related to a relaxation time $\tau_{r}\rightarrow 0$ at $d=d_{c}$ and a corresponding small parameter.
 This result also shows that the phenomenon is very robust:
all one needs is a point at which energy cascade changes its direction. The results of the shell model investigation will be published elsewhere [21].

\noindent The expression
(47) is similar to the one obtained in a  ground-breaking 
paper by Polyakov on the scale-invariance of strong interactions,  where the 
 multifractal scaling and the pdf were 
 analytically derived for the first time [22]-[23]. In the review paper 
[23] Polyakov noticed
that the exact result can be simply reproduced considering a cascade process 
with a heavy stream (particle) transformed into lighter streams at each step
of the cascade (fission).  
Due to   the relativistic effects the higher the energy of the
particle, the smaller  the angle of a cone,  accessible to the
fragments formed as a result of  fission. Thus, the larger
 the number of a cascade step, the smaller is the
fraction of space occupied by the particles [23]. 

\noindent The theory, presented in this paper describes many experimental
observations. Still,  understanding of  the limits of validity of  
expression
(29) is crucial for the final assessment of the theory. The relation (32) 
shows
that (29) is consistent with the gaussian tails of the pdf. However, at the
present stage we are
unable to prove that (29) is the only expression leading to this result.  The
problem is that without experimental detection of  at least 
some deviations from the 
the gaussian statistics of transverse velocity differences, 
 one will not be able
to understand the limits of validity of (29). Given the state-of-the-art of
numerical simulations, this goal may not be that simple.

\begin{figure}
\centerline{\psfig{file=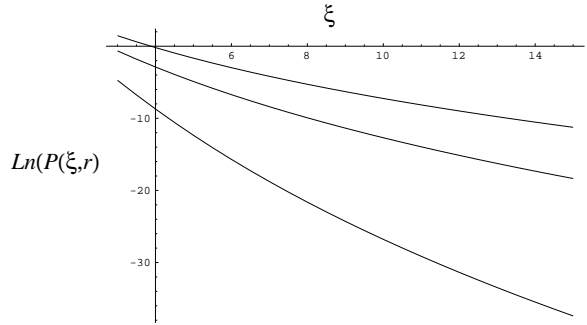,width=3.0in}}
%\vspace*{-3.0in}
\caption{$Ln(P(\xi,r))$.~ 
From bottom to top:  r/L=0.1;~0.01;~0.001, respectively.}
\label{fig.pdf}
\end{figure}

\section{ Acknowledgment} I am grateful to A. Polyakov for most interesting
and stimulating discussions. Help from K.R. Sreenivasan 
and M. Vergassola,  who shared with me their sometimes unpublished experimental
data , was most usefull. Recent numerical results  of M. Jensen produced not
 only support of  conclusions of this paper but also demonstrated  a surprising 
power and importance of a  shell model.
I benefited a lot from conversations with B. Shraiman, 
P. Constantin, C. Doering,
T. Gotoh and T. Kambe.  

\section{References}
1. M. Cherkov, G. Falkovich, I. Kolokolov and V. Lebedev, Phys. Rev. E, {\bf 52}, 4924 (1995)\\
2. K. Gawedzki and A. kupiainen, Phys. Rev. Lett., {\bf 75}, 3834 (1995)\\
3. B. I. Shraiman and E. D. Siggia, C.R. Acad. Sci., {\bf 321}, Serie II, 279 (1995)\\
4. L.Smith and V. Yakhot, J. Fluid. Mech. {\bf 274},  115 (1994)\\
5. L.Smith and V. Yakhot, Phys.Rev.Lett. {\bf 71}, 352 (1993)\\
6. P. Tabeling and J. Paret, Phys. Fluids, {\bf 12}, 3126 (1998)\\
7. G. Bofetta, A. Cellini, M. Vergassola, Chao-Dyn/9906016\\
8. U. Frisch, ``Turbulence'', Cambridge University Press, 1995\\
9. U. Frisch and J.D. Fournier, Phys. Rev. A, {\bf 17}, 747 (1978)\\
10. V. Yakhot and S.A.Orszag, Phys.Rev.Lett. {\bf 57}, 1722 (1986)\\
11. V. Yakhot, Phys. Rev.E, in press (1999)\\
12. A.M. Polyakov, Phys.Rev. E, {\bf 52}, 6183 (1995)\\
13. L.D.Landau and E.M. Lifshitz, Fluid Mechanics, Pergamon Press, Oxford, 198
14. A.S.Monin and A.M.Yaglom, ``Statistical Fluid Mechanics'' vol. 1, 
MIT Press,
Cambridge, MA (1971)\\
15. Ya. G. Sinai and V. Yakhot, Phys.Rev.Lett., {\bf 63}, 1962 (1989)
\\
16. R.H.Kraichnan, Phys.Fluids. {\bf 10}, 1417 (1967)\\
17. Yakhot Orszag, J. Sci. Comp., {\bf 1}, 3 (1986)\\
18. V. Yakhot, Phyds. Rev. E., {\bf 55}, 329 (1997)\\
~ V. Yakhot, Phys. Rev.E,  Phys. Rev. E, {\bf 57}, 1737 (1998)\\
19. K.R. Sreenivasan, private communication\\
20. M. Jensen, private communication\\
21. M. Jensen and V. Yakhot, to be published\\
22. A.M. Polyakov, Sov. Phys. JETP {\bf 34}, 1177 (1972)\\
23. A.M. Polyakov, ``Scale Invariance of Strong Interactions and Its
    Application to Lepton-Hadron Reactions'', preprint,
 Landau Institute for Theoretical Physics, 1971\\

\end{document}